\documentclass[prb,twocolumn,showpacs,amsmath]{revtex4}

\usepackage{graphicx}
\usepackage{bm}
\usepackage{times}

\newcommand{\smeq}{\! = \!}

\newcommand{\smpl}{\! + \!}
\newcommand{\smmi}{\! - \!}

\newcommand{\ve}{\varepsilon}

\newcommand{\Ef}{E_{\text{F}}}

\newcommand{\ag}{{\mathcal{N}}}
\newcommand{\kt}{k_{\text{B}}T}

\newcommand{\be}{\begin{equation}}
\newcommand{\ee}{\end{equation}}
\newcommand{\bea}{\begin{eqnarray}}
\newcommand{\eea}{\end{eqnarray}}
\newcommand{\Ha}{{\hat H}}

\newcommand{\up}{\uparrow}
\newcommand{\dn}{\downarrow}

\begin{document}
\title{Exchange and the Coulomb blockade: Peak height statistics in quantum dots}
\author{Gonzalo Usaj}
\author{Harold U. Baranger}
\affiliation{Department of Physics, Duke University, P.O.Box 90305,
  Durham North Carolina 27708-0305}
\date{Received 29 Nov 2002}

   \begin{abstract}
We study the effect of the exchange interaction on the Coulomb
blockade peak height statistics in chaotic quantum dots. 
Because exchange reduces the level repulsion in the many body spectrum, it
strongly affects the fluctuations of the peak conductance at finite
temperature. We find that including exchange substantially
improves the description of the experimental data. Moreover, it
provides further evidence of the
presence of high spin states ($S\ge1$) in such  systems. 
   \end{abstract}
\pacs{73.23.Hk, 73.40.Gk, 73.63.Kv}

\maketitle 

Transport properties of quantum dots (QDs) are strongly
affected by interactions. A paradigmatic example is the Coulomb blockade
(CB) of electron tunneling.
\cite{KouwenetalRev97,Alhassid00,AleinerBG02}
It occurs at low temperature when
the thermal energy is smaller than the charging energy ($E_{C}$)
required to add an electron to the QD. The conductance is then 
blockaded, being restored only at specific values of
the gate voltage $V_{g}$ of a nearby gate capacitively coupled to the
QD. This leads to a series of sharp CB conductance peaks. 
The statistical properties of these peaks reflect the
mesoscopic fluctuations of the (many-body) spectrum and wave-functions
of the QD. In particular, they encode information about the
non-trivial spin statistics that arises from the presence of exchange, 
the most important part of the residual interaction.\cite{AleinerG98,AleinerBG02}   

Both the peak height distribution (PHD) and the peak spacing distribution
 have been studied in detail (see Ref. \onlinecite{Alhassid00}
for a review), though the latter has received most of the attention. 
This is mainly due to the fact that early
 measurements\cite{ChangBPWC96,FolkPGHCM96} of the PHD were found to
 be in good agreement with the constant interaction (CI)
 model,\cite{JalabertSA92,AlhassidW02} contrary to the case of the  peak
 spacing. In this simple model of CB, the \textit{e-e} interaction is
 assumed constant (given by $E_C$) and the fluctuations of the single
 particle properties described by random matrix theory---appropriate
 for chaotic (or diffusive) QDs.

A later experiment\cite{PatelSMGASDH98} showed, however, significant
deviations from the CI model. Namely, the peak height
fluctuations were found to be smaller than expected in the entire experimental
temperature range ($\kt\smeq0.1\!-\!2\Delta$, with $\Delta$ the single
particle mean level spacing).
The authors attributed the discrepancy at high $T$ to the presence
of dephasing but the origin of the low $T$ behavior was
unclear---a recent calculation\cite{RuppAM02} showed
that dephasing by itself cannot account for the high $T$ data
either. It was then proposed\cite{HeldEA02_SO} that the observed
reduction of the fluctuations at low $T$ is related to the presence of
spin-orbit coupling. Though the result is in good agreement with the
low $T$ data, the assumed strength of the spin-orbit interaction
requires some test---for instance, experimental data in
Ref. \onlinecite{ZumnuhlMMCG02} suggest that in small QDs (as used in
Ref. \onlinecite{PatelSMGASDH98}) the effect of spin-orbit is rather weak.
 
So far, however, exchange---the main interaction  effect---has not been
 considered.
This contrasts with the case of the peak spacing distribution where
it was shown to be crucial, as it leads to the appearance of
non-trivial spin states in the QD
($S\!\ge\!1$)\cite{BrouwerOH99,BarangerUG00,KurlandAA00,UllmoB01} and
enhances the effect of finite temperature.\cite{UsajB01_RC,UsajB02}  

In this work we show that exchange also strongly affects the PHD. Furthermore,
it accounts for most of the
disagreement with the low $T$ data of Ref. \onlinecite{PatelSMGASDH98},
\textit{without} including dephasing or spin-orbit coupling. Such agreement
is yet another indication of the presence of high  spin states in QDs. 

A chaotic QD containing a large number of electrons can be described
by the following Hamiltonian\cite{AleinerG98,KurlandAA00,AleinerBG02} 
\be
\Ha_{}^{}\smeq\sum_{\alpha,\sigma}{\ve_{\alpha}}\,\hat{n}_{\alpha,\sigma}
                              \smpl E_{C}\,(\hat{n}-\ag)^2\smmi J_{S}\;{\vec S}^2
\label{CEI}
\ee
where $\{\ve_{\alpha}\}$, the single particle energies, are described
by random matrix theory, $\ag\smeq
C_g V_g/e$ describes the capacitive coupling to the control gate, $C_g$
is the dot-gate capacitance, ${\vec S}$ is the total spin operator, and
$J_{S}$ is the exchange constant.
The second and third terms in Eq. (\ref{CEI}) describe the QD charging
energy and the exchange interaction, respectively. 
We assume elastic transport and broken time-reversal
symmetry---therefore we use the Gaussian Unitary Ensemble
(GUE)---unless otherwise stated. 

In the regime $\Gamma\ll \kt,\Delta\ll E_C$, where $\Gamma$ is the
total width of a level in the QD, the conductance near the CB
peak corresponding to the $N\smmi1\!\rightarrow\!N$ transition is given by
\cite{Beenakker91,MeirW92,Pastawski92,UsajB02} 
\bea
\nonumber
G(x)&\smeq&\frac{e^2}{\hbar \kt}\,
        \sum_{i,j,\alpha,\sigma}{\frac{\Gamma_{\alpha}^L\Gamma_{\alpha}^R}
         {\Gamma_{\alpha}^L\smpl\Gamma_{\alpha}^R}\,
 \left|\langle\Psi_{j}^{N}|c_{\alpha,\sigma}^{\dagger}|\Psi_{i}^{N\smmi1}\rangle\right|^2}\\
&&\times
\frac{F_\text{eq}(j)F_\text{eq}(i)}{\left[\sqrt{F_\text{eq}(j)}\smpl\sqrt{F_\text{eq}(i)}
\,\right]^{2}}\,g_{ji}^{}(x\smmi x^{*}_{ji})
\label{G}
\eea
where $x\smeq2E_{C}[\ag\smmi (N\smmi\frac{1}{2})]$ and $x^{*}_{ji}$ is
defined below.
Here, (i) $\Gamma^{L(R)}_{\alpha}$ is the partial width of the
single-particle level $\alpha$ due to tunneling to the left
(right) lead, (ii) $\{|\Psi_{j}^N\rangle\}$ are the eigenstates of the
QD with $N$ electrons, (iii) $F_\text{eq}(j)$ is the
canonical probability that the eigenstate $j$ is occupied, and (iv)
\be 
g_{ji}^{}(x)\smeq \frac{f(y_{ji}^{}\smmi x)f(y_{ji}^{}\smpl
x)}{f(y_{ji}^{})^{2}}
\ee
with $f(x)\smeq(1\smpl\exp[x/\kt])^{\smmi1}$ and 
$y_{ji}^{}\smeq\kt\ln\left[F_\text{eq}(j)/F_\text{eq}(i)\right]$. 
Notice that $g_{ji}(x)$ reaches its maximum value at $x\smeq0$. 
Let us denote by $\{E_{j}^{}\}$ the eigenvalues of $\Ha_\text{}$
without the charging energy term. Then, the contribution of the
transition $i\!\rightarrow\!j$ to the conductance reaches 
its maximum when 
$x\smeq x_{ji}^{*}\!\equiv\! E_{j}^{N}\smmi E_{i}^{N\smmi1}\smpl
y_{ji}^{}/2-\Ef$. Note the shift in the peak 
position due to finite temperature.\cite{GlazmanM88,Akera99a,UsajB01_RC,UsajB02,Beenakker91}
Since both $F_\text{eq}$ and the overlap 
$\left|\langle\Psi_{j}^{N}|c_{\alpha,\sigma}^{\dagger}|\Psi_{i}^{N\smmi1}\rangle\right|^2$
depend on the spin of the two states involved in the transition, it is
clear that this maximum is spin-dependent.\cite{GlazmanM88,Beenakker91,Akera99a,WeinmannHK95} 
 
The eigenstates of the Hamiltonian (\ref{CEI}) can be written as
$|\Psi_{j}^N\rangle\smeq|\{n'_{\alpha}\},S',S_{z}',k'\rangle$, with 
$n'_{\alpha}\smeq n'_{\alpha,\up}\smpl n'_{\alpha,\dn}$ the occupation
of the $\alpha$-th single particle state.\footnote{An extra quantum
  number $k'$ is required when the number of singly occupied levels is
  bigger than $2$ (see Ref. \onlinecite{Pauncz}).} Since there is no
energy dependence on $S_{z}'$---Zeeman energy is neglected---the sum
over $S_z$, $S_{z}'$ and $\sigma$ in Eq. (\ref{G}) can be easily
carried out. One then gets\cite{UsajB02,Pauncz}
\be
\sum_{S_{z}',k' \atop S_{z},k,\sigma}{\left|\langle\Psi_{S_z',k'}^{N}|c_{\alpha,\sigma}^{\dagger}
|\Psi_{S_z,k}^{N\smmi1}\rangle\right|^2}\smeq\left\{\!
\begin{array}{ll}
\!(2S'\smpl1)N_{k}& \text{if $n_{\alpha}\smeq0$}\\
&\\
\!(2S\smpl1)N_{k'}&\text{if $n_{\alpha}\smeq1$}\\
\end{array}
\right.
\label{overlap}
\ee
where $n_{\alpha}$ refers to the state with $N\smmi1$
 particles, $N_k$ ($N_{k'}$) is  the degeneracy associated with $k$
 ($k'$)\footnote{If $N_S$ is the number of singly occupied levels,
 then $N_k\smeq\left({N_S \atop
 \frac{1}{2}N_{S}-S}\right)\smmi\left({N_S \atop
 \frac{1}{2}N_{S}-S-1}\right)$.} and $|S'\smmi S|\smeq\frac{1}{2}$.
\begin{figure}[t]
\begin{center}
 \includegraphics[width=8.5cm,clip] {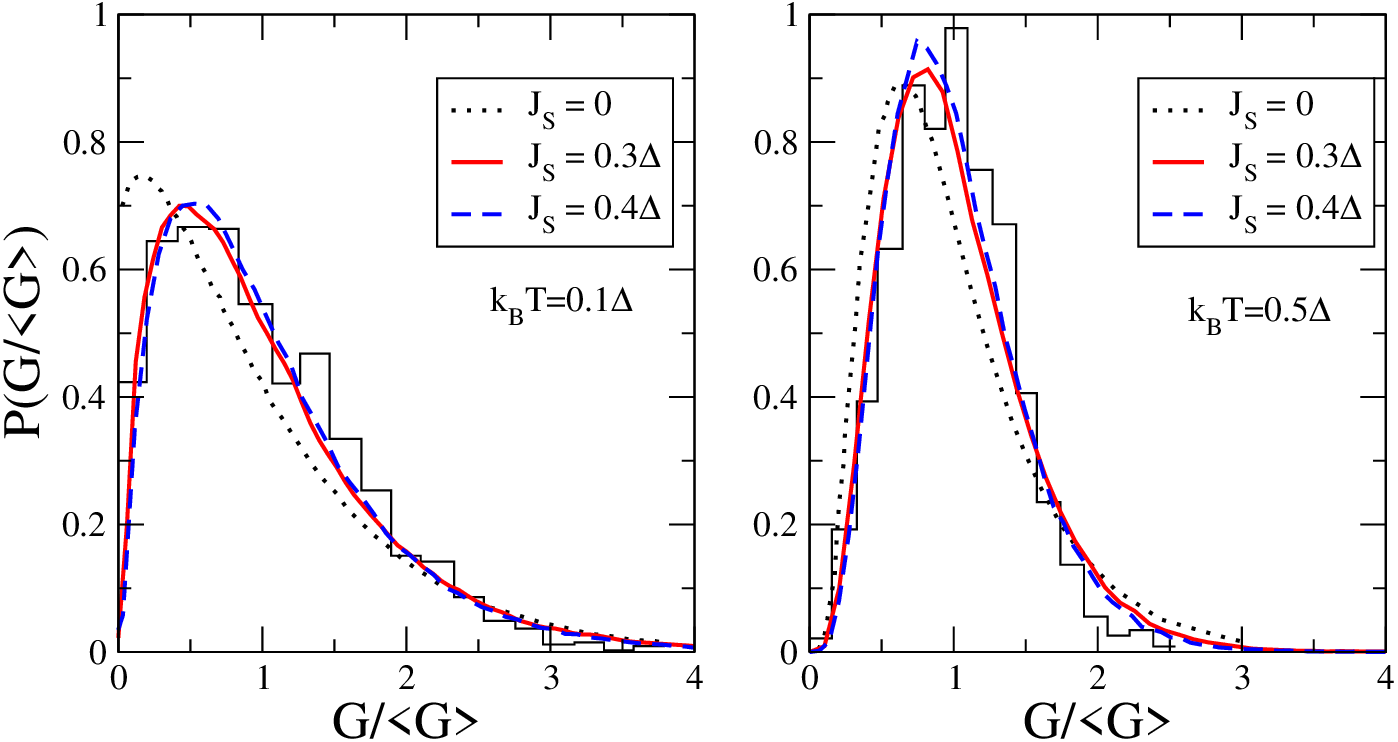}
\end{center}
\caption{Coulomb blockade peak height distribution at finite
  temperature. The theoretical distributions (lines) were 
  obtained by numerically solving Eq. (\ref{G}) for different values
  of the exchange constant ($J_{S}$). The histograms correspond to the
  experimental data in Ref. \onlinecite{PatelSMGASDH98}. 
Notice that good agreement is obtained only after the addition of
  exchange.
}
\label{distribution}
\end{figure}

In the special case of very low $T$, when only the ground state is relevant, we have
$F_\text{eq}(j)\!\simeq\!1/(2S'\smpl1)$ ,
$F_\text{eq}(i)\!\simeq\!1/(2S\smpl1)$, and
$N_{k'}\smeq N_{k}\smeq1$, so that the peak conductance is given by 
\be
G_{\text{peak}}\smeq\lambda_{S',S}\frac{2e^2}{\hbar \kt}\, 
\frac{\Gamma_{\alpha}^L\Gamma_{\alpha}^R}{\Gamma_{\alpha}^L\smpl\Gamma_{\alpha}^R}
\label{maxconduct}
\ee
with
\be
\lambda_{S',S}\smeq  \frac{2(S'\smpl S)\smpl 3}
              {4\,\left(\sqrt{2S'\smpl1}\smpl\sqrt{2S\smpl1}\right)^{2}}
\ee
where still $|S'\smmi S|\smeq\frac{1}{2}$.
Because the probability for the $S\!\rightarrow\!S'$ transition
depends on both the interaction strength ($J_{S}$) and the 
statistics of the single particle spectrum, it turns out that the PHD
depends, even at low $T$, on the fluctuations of \textit{both} the
wave-functions ($\Gamma$'s) \textit{and} the many-body spectrum. 

At finite temperature many transitions contribute to the
conductance. We calculate the PHD by finding the maximum of
Eq. (\ref{G}) numerically. According to random matrix theory, we describe the
fluctuation of the widths $\Gamma^{q}_{\alpha}$ with $q\smeq L,R$ by the
Porter-Thomas distribution
$P(\Gamma^{q}_{\alpha})\smeq\bar{\Gamma}^{-1}\exp(\smmi\Gamma^{q}_{\alpha}
/\bar{\Gamma})$, appropriate for the GUE.  We keep states within an energy window of
$\max(2\Delta, 6\kt)$ around the ground state.
\begin{figure}[t]
\begin{center}
 \includegraphics[width=7.5cm,clip] {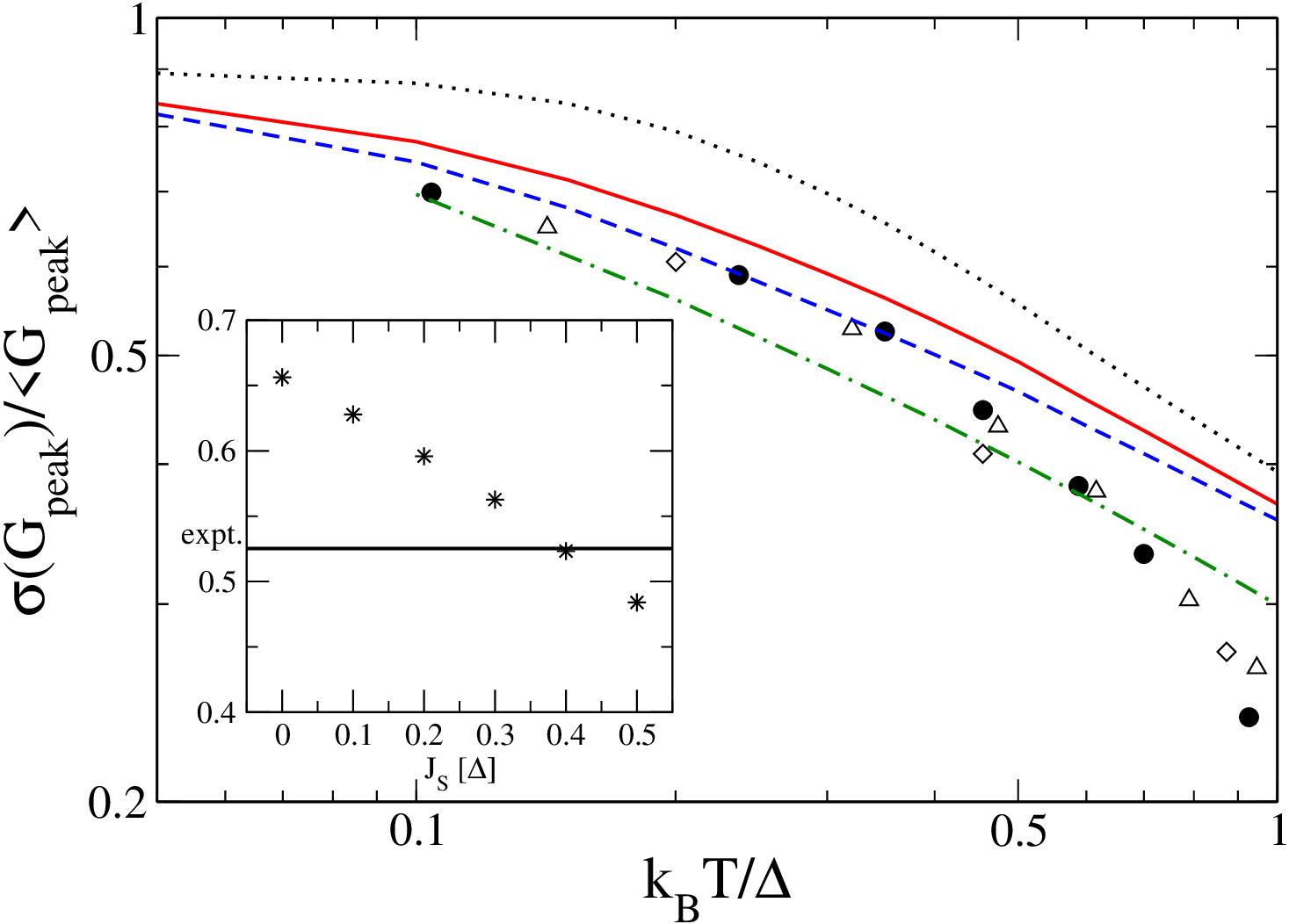}
\end{center}
\caption{Ratio of the root mean square to the mean of the conductance
peak as a function of temperature. Different curves correspond to
different values of exchange $J_{S}\smeq0$, $0.3\Delta$, and
$0.4\Delta$ (dotted, solid, and dashed lines, respectively). Symbols
correspond to the data in Ref. \onlinecite{PatelSMGASDH98}. Notice the
strong reduction of the fluctuation introduced by the exchange
interaction at low temperature. The inset shows the decay of
$\sigma(G_{\text{peak}})/\langle G_{\text{peak}}\rangle$ as a function
of $J_S$ at $\kt\smeq0.35\Delta$.
At high $T$ the experimental data show a stronger suppression of the
fluctuations than predicted in the strong inelastic regime for
$J_S\smeq0.4\Delta$ (dot-dashed line). 
}
\label{sigma}
\end{figure}
 
Figure \ref{distribution} shows the PHD for $\kt\smeq0.1\Delta$,
$0.5\Delta$ and $J_{S}\smeq0$, $0.3\Delta$, $0.4\Delta$. The histograms
correspond to the experimental data in Ref. \onlinecite{PatelSMGASDH98}. 
At low temperature the agreement is very good for the non-zero 
values of the exchange constant but clearly not for $J_S\smeq0$ (CI model). 
This is indirect evidence for the presence of high
spin states in QDs---we should point out though that spin-orbit leads to
a similar effect if it is strong enough.\cite{HeldEA02_SO}
At higher temperature the presence of exchange improves the fit
but not enough to fully account for the observed distribution. 
Nevertheless, it is clear that exchange substantially modifies the PHD
and thus cannot be ignored.

Since an accurate measurement of the full PHD is quite demanding, a
detailed comparison with theory is difficult. Instead, it is usually more
convenient to look at the first few moments.
In Ref. \onlinecite{PatelSMGASDH98}, the 
ratio of the root mean square and the mean value of the conductance peak height,
$\sigma(G_{\text{peak}})/\langle G_{\text{peak}}\rangle$, was measured
as a function of temperature. It was found to be smaller than the value predicted by
the CI model.   
Figure \ref{sigma} compares the experimental data with our results for
different values of $J_S$. The inset shows the $J_S$-dependence at a fixed
temperature ($\kt\smeq0.35\Delta$).

The improvement introduced by the exchange interaction at low
temperature is evident. This can be easily understood as follows:
because exchange reduces the level repulsion in
the \textit{many body} spectrum, the number of levels that contribute to the
conductance at a given temperature increases and therefore the
fluctuations are reduced---in fact, the importance of the interplay between
temperature and exchange was pointed out in
Ref. \onlinecite{UsajB01_RC}. This is similar to the  mechanism
discussed in Ref. \onlinecite{HeldEA02_SO} where spin-orbit coupling
reduces the repulsion in the \textit{single particle} spectrum.
At very low temperature, Eq. (\ref{maxconduct}) 
predicts an enhancement of the fluctuations compared to the CI
model result, $\sigma(G_{\text{peak}})/\langle
G_{\text{peak}}\rangle\smeq [\langle \lambda_{S',S}^2\rangle/\langle
\lambda_{S',S}\rangle^2\times 9/5\smmi1]^{1/2}\!\ge\!2/\sqrt{5}$. 
However, this enhancement is negligible since 
$\langle \lambda_{S',S}^2\rangle/\langle \lambda_{S',S}\rangle^2\simeq1$ 
for typical values of the interaction.
We also checked that the effect on Fig. \ref{sigma} of non-universal
corrections\cite{AleinerBG02} to the Hamiltonian (\ref{CEI}) is very
small, consistent with the results for the peak spacing
distribution where they mainly affect the shape of the
distribution.\cite{UsajB01_RC,UsajB02,AlhassidM02}
 
Our model shows a significant deviation from the experimental data for
$\kt\!\ge\!0.5\Delta$. It is tempting to attribute this to the presence
of inelastic processes, which we have not taken into account so
far. Notice that after exchange is included, dephasing only needs to
account for a (small) fraction of the reduction of the
fluctuations. 
Accounting for an arbitrary inelastic rate requires solving a master
equation for the transition probabilities, which is not a simple
task.\cite{Beenakker91,EisenbergHA02} Instead, we calculate the peak
conductance in the strong inelastic regime,\cite{BeenakkerSS01} where
electrons inside the QD are assumed to be in thermal equilibrium. 
In practice, this means that instead of calculating the thermal
average of
$\Gamma_{\alpha}^L\Gamma_{\alpha}^R/(\Gamma_{\alpha}^L\smpl\Gamma_{\alpha}^R)$
---see Eq. (\ref{G})---, one first calculates the thermal average of
the couplings $\{\Gamma^q_{\alpha}\}$ and then the ratio,
$\langle\Gamma^L\rangle\langle\Gamma^R\rangle/\langle\Gamma^L\smpl\Gamma^R\rangle$.
This is expected to be a lower bound for the experimental data. Figure
\ref{sigma} shows that this is \textit{not} the case. The reason for
this is unclear.
\begin{figure}[t]
\begin{center}
 \includegraphics[width=7.5cm,clip] {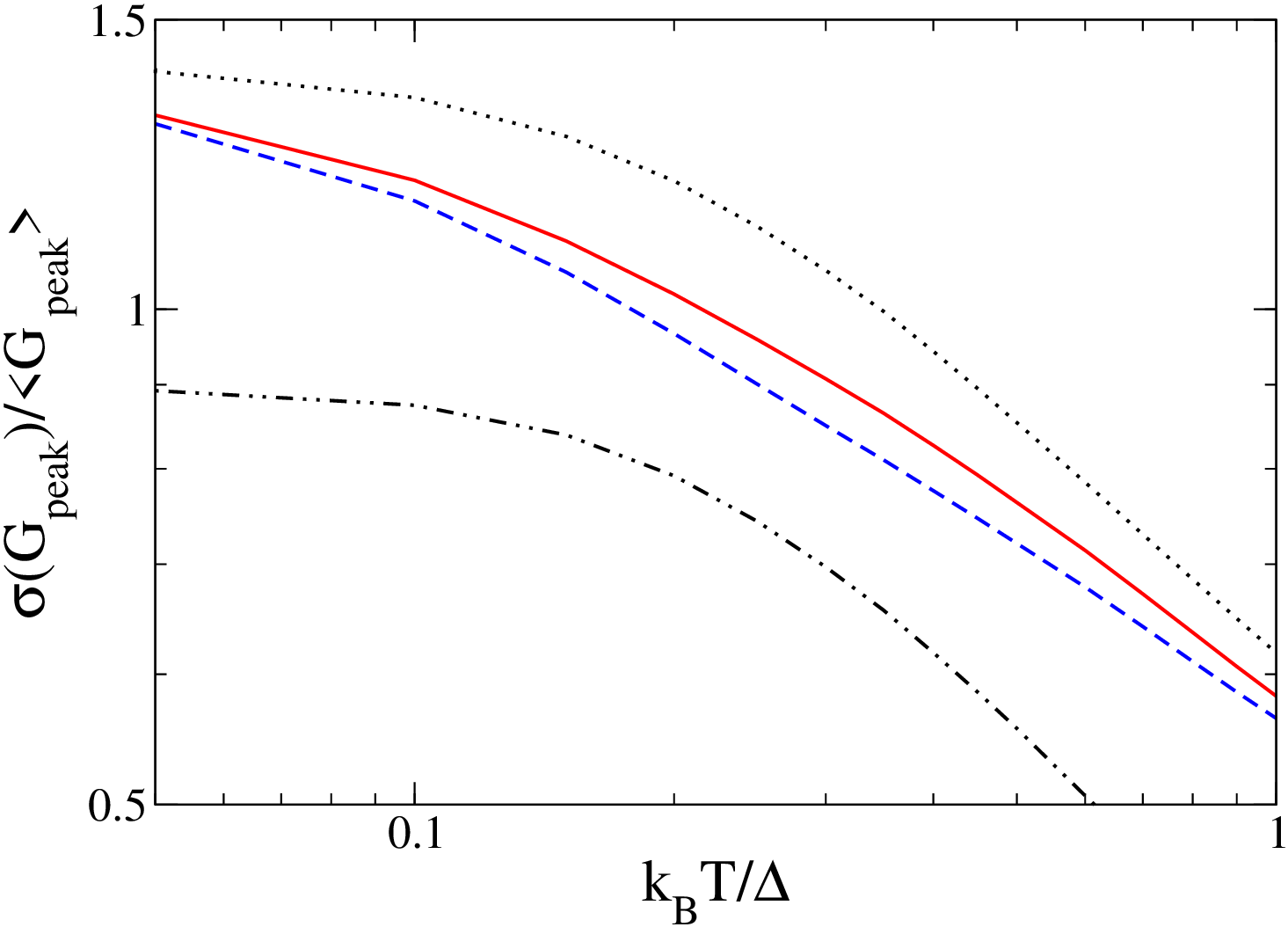}
\end{center}
\caption{Same as Figure \ref{sigma} but for zero magnetic field
  (GOE). The GUE result for $J_{S}\smeq0$ is included for comparison
  (dot-dot-dash line).
}
\label{sigmaGOE}
\end{figure}

A direct comparison between our results and the experimental data seems to
suggest that $J_S\!\simeq\!0.4\Delta$ (see inset in Figure
\ref{sigma}). This value of the interaction is
bigger than what one would estimate from an RPA calculation of the
screened potential for the experimental density
($J_S\!\simeq\!0.3\Delta$).\cite{AleinerBG02,UsajB02} At
present it is not clear to us whether
 corrections beyond RPA are required,\cite{JiangBW02,OregBWH01}
or if other effects such as spin-orbit\cite{HeldEA02_SO} need to
be included.

It is interesting, then, to consider ways to distinguish
between the exchange and spin-orbit scenarios. In the
latter case, it was predicted\cite{HeldEA02_SO} that an in-plane
magnetic field will restore the level repulsion and increase the
fluctuations. The behavior is the opposite in the absence of a
perpendicular magnetic field, where the system will evolve (as the
in-plane field is increased) from the orthogonal to the
unitary ensemble (see Figure \ref{sigmaGOE}).\cite{HeldEA02_SO} 
The same general trend is also valid in the exchange-dominated scenario
because a large enough parallel field will always drive the system
towards the strong spin-orbit regime.\cite{FolkPBMDH01} Therefore, the
only difference between the two scenarios would be
the specific dependence of  $\sigma(G_{\text{peak}})/\langle
G_{\text{peak}}\rangle$ on the magnitude of the parallel field.
\begin{figure}[t]
\begin{center}
 \includegraphics[width=7.2cm,clip] {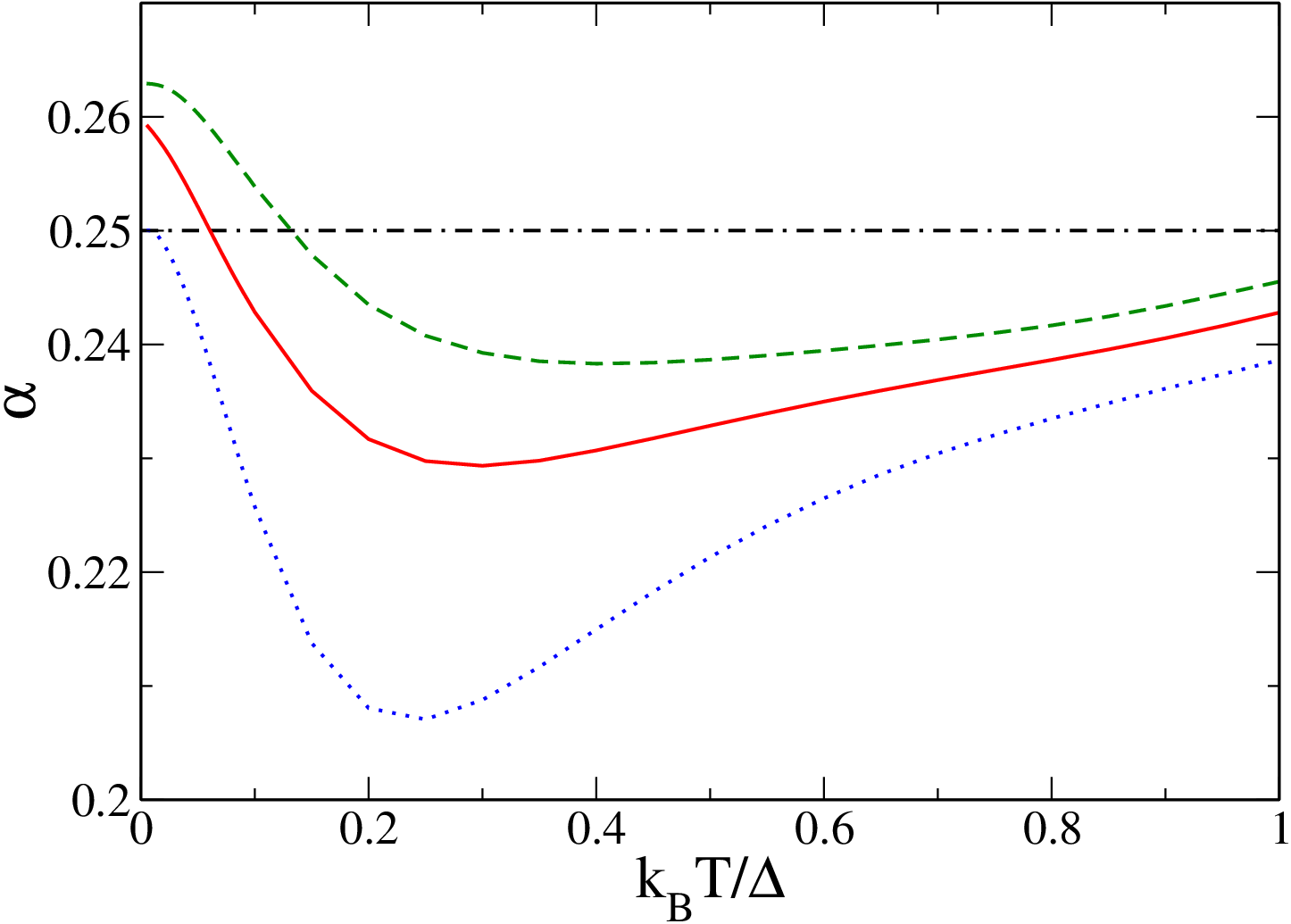}
\end{center}
\caption{Relative change in the mean conductance upon breaking
 time-reversal symmetry [$\alpha$, see Eq. (\ref{alpha})] as a function of
 temperature. Different curves correspond to different values of the
 exchange constant: $0$ (dotted), $0.3\Delta$ (solid), and $0.4\Delta$
 (dashed).
Notice that $\alpha$ is increased by exchange and, in particular, that
it is bigger than $0.25$ for 
$T\!\rightarrow\!0$ when $J_S>0$.
}
\label{alpha}
\end{figure}

A sharper distinction between the two scenarios could be made by
changing the electron density $n_e$ in the QD. On the one hand, an
increase of $n_e$ will slightly decrease $J_S$ and therefore enhance
the fluctuations (see Figures \ref{sigma} and \ref{sigmaGOE}). On the
other hand, increasing $n_e$ increases the magnitude of
the spin-orbit coupling\cite{MillerZMLGCG02,ZumnuhlMMCG02} and so
decreases the conductance fluctuations---according to
Ref. \onlinecite{ZumnuhlMMCG02}, we assume that the initial value of
the spin-orbit coupling is smaller than the one required to complete
the crossover described in Ref. \onlinecite{HeldEA02_SO}.  Therefore,
the sign of $\partial [\sigma(G_{\text{peak}})/\langle
G_{\text{peak}}\rangle]/\partial n_e$ at low $T$ is different for each
scenario---a sharp distinction. 

It is worth pointing out that an independent measurement of
 $J_S$---obtained, for instance, by measuring the spin
 distribution\cite{JacquodS00,KurlandBA01,FolkMBKAA01,LindemannIHZEMG02}---would be a direct
 way to rule out or quantify the effect of spin-orbit coupling and
 exchange.
 
Finally, we study the effect of exchange on  the relative change of
the mean value of the conductance upon breaking time-reversal symmetry,
\be
\alpha\!\smeq\!1\smmi\frac{\langle G_{\text{peak}}\rangle_{\text{GOE}}}{\langle
G_{\text{peak}}\rangle_{\text{GUE}}}.
\ee 
Very recently, the temperature dependence of this quantity was measured
by Folk \textit{et al}.,\cite{FolkMH01} who used it to estimate the
dephasing time ($\tau_{\phi}$) in weakly coupled
QDs.\cite{AltshulerGKL97}
This estimate is based on the deviation of the experimental value
of $\alpha$ from the one expected within the CI model in the elastic
regime, namely, $\alpha\!\simeq\!0.25$ for both $\kt\!\ll\!\Delta$
and $\kt\!\gg\!\Delta$.\cite{Alhassid98} For intermediate
temperatures $\alpha$ is slightly smaller than $0.25$ because of the difference in
the spectral fluctuations of the two ensembles.\cite{RuppAM02,HeldEA02}

Figure \ref{alpha} illustrates the temperature dependence of $\alpha$
for three different values of $J_{S}$. 
Notice that $\alpha$ increases monotonically with $J_S$
in the full temperature range.
That is, exchange enhances $\alpha$.
In particular, the bound
$\alpha\!\le\!0.25$ is no longer valid  at low
temperature.\cite{UsajB02} This can be verified using
Eq. (\ref{maxconduct}),  
\be
\alpha\!\simeq\!1\smmi\frac{3}{4}\frac{\langle
  \lambda_{S',S} \rangle_{\text{GOE}}}{\langle\lambda_{S',S}\rangle_{\text{GUE}}}
\!>\!0.25\,.
\ee
This enhancement originates in the difference between the
spin distributions in the two ensembles---high spin states are more
likely to occur in the GOE. 
Also note that the effect of the spectral
fluctuations is substantially reduced.
As a result, $\alpha$ remains closer to $0.25$ than in the CI
model. Unfortunately, because of large statistical errors in the
experimental data,\cite{FolkMH01} this difference could not be discerned.

Since inelastic processes reduce $\alpha$,\cite{BeenakkerSS01,EisenbergHA02}
the bigger the difference between the elastic result and
the measured value of $\alpha$ the shorter $\tau_{\phi}$. 
Then, it is clear that neglecting exchange
overestimates $\tau_{\phi}$. Though we do not expect the difference
to be large, it could be relevant when deciding whether or not there is
a saturation of $\tau_{\phi}$. 

In summary, we have studied the effect of the exchange interaction on the
peak height distribution. We found that it strongly affects the
distribution at finite temperature.
Our results indicate that the experimental data present signatures of
high spin states and suggest that $J_S\!\gtrsim\!0.3\Delta $ for  the dots of Ref. 
\onlinecite{PatelSMGASDH98}.

We appreciate helpful discussions with K. Held,  E. Eisenberg,
P. Brouwer, and  B. L. Altshuler. GU acknowledges partial support from
CONICET (Argentina). This work was supported in part by the NSF
(DMR-0103003). After submission of this paper, we learned of
Ref. \onlinecite{AlhassidR02} which contains some overlapping material.

\bibliographystyle{apsrev}

\end{document}